\documentclass[sigconf]{acmart}

\copyrightyear{2021} 
\acmYear{2021} 
\setcopyright{acmcopyright}\acmConference[ICAIF'21]{2nd ACM International Conference on AI in Finance}{November 3--5, 2021}{Virtual Event, USA}
\acmBooktitle{2nd ACM International Conference on AI in Finance (ICAIF'21), November 3--5, 2021, Virtual Event, USA}
\acmPrice{15.00}
\acmDOI{10.1145/3490354.3494369}
\acmISBN{978-1-4503-9148-1/21/11}

\usepackage{xcolor}
\usepackage{svg}
\usepackage{float}
\usepackage{caption}
\usepackage{subcaption}
\usepackage{enumitem}
\setlist[enumerate]{nosep}
\begin{document}
\setlength{\abovedisplayskip}{3pt}
\setlength{\belowdisplayskip}{3pt}
\settopmatter{printacmref=false}

\title{Profit equitably: An investigation of market maker's impact on equitable outcomes}

\author{Kshama Dwarakanath, Svitlana S Vyetrenko and Tucker Balch}
\email{{kshama.dwarakanath,svitlana.s.vyetrenko}@jpmchase.com}
\affiliation{%
  \institution{JP Morgan AI Research, USA}\country{}
}

%
\begin{abstract}
{
We look at discovering the impact of market microstructure
on equitability for market participants
at public exchanges such as the New York
Stock Exchange or NASDAQ. Are these environments
equitable venues for low-frequency participants
(such as retail investors)? In particular,
can market makers contribute to equitability for
these agents? We use a simulator to assess the effect a market marker can have on equality of outcomes for consumer or retail traders by adjusting its parameters. Upon numerically quantifying market equitability by the entropy of the price returns distribution of consumer agents, we demonstrate that market makers indeed support equitability and that a negative correlation is observed between the profits of the market maker and equitability. We then use multi objective reinforcement learning to concurrently optimize for the two objectives of consumer agent equitability and market maker profitability, which leads us to learn policies that facilitate lower market volatility and tighter spreads for comparable profit levels.
}
\end{abstract}

\keywords{Agent based simulations, equitability, multi objective reinforcement learning, volatility}


\maketitle

\section{Background and related work}

\subsection{Introduction}
{

An estimated 40\% to 60\% of trading activity across all financial markets (stocks, derivatives, liquid foreign currencies) is due to high frequency trading, which is when trading happens electronically at ultra high speeds \cite{kirilenko_lo}. High frequency trading and its implications on market stability and equitability have been in regulatory spotlight after the flash crash of May 6, 2010, when the Dow Jones Industrial Average dropped 9\% within minutes \cite{kirilenko_flash_crash}.
To dissect the impact of high frequency trading, it is necessary to distinguish between intentional manipulation (e.g. front-running, spoofing) and legitimate trading strategies such as market making, pairs trading and statistical arbitrage, new reaction strategies, etc \cite{high_freq_fairness}. Designated market makers (denoted MM henceforth), for example, are obliged to continuously provide liquidity to both buyers and sellers regardless of market conditions - contributing to market stability by providing counterparty to all transactions even in times of market distress. For example, NASDAQ requires MMs "to maintain a continuous two-sided trading interest during regular market hours, at prices within certain parameters expressed as a percentage referenced from the National Best Bid or Offer" \cite{NASDAQ_SEC}. MMs are typically rewarded with lower exchange fees for doing so \cite{Foucault_fees}. 

MM presence and facilitation of transactions among market participants is associated with better market quality in terms of high frequency of auction clearance, and low variability in returns and trading volume \cite{mm}. Classical approaches to MM include modeling the order arrivals and executions, and using control algorithms to find an optimal strategy for the MM \cite{chakraborty2011market}. Reinforcement learning (RL) has also been used to learn market making strategies using a replay of historical data to derive MM's policies \cite{sutton2018reinforcement,SpoonerMarketMaking,chan2001electronic}.

Replaying historical data in MM agent training does not take into account trading response from the market environment. In this paper, we study how to learn a MM strategy that optimizes for both equitability and profitability in a responsive agent-based simulated environment. Specifically, }we examine market equitability in scenarios where an elementary MM provides liquidity in an equity market including a large number of consumer or retail investors. The MM is defined by a set of parameters that govern the spread and depth of orders it places into the order book. Based on observations of the relationship between the MM profits and market equitability with variations in the MM parameters, we formulate and solve an RL problem to find a MM policy that optimizes for both market equitability as well as MM profits. The contributions of this paper are as follows
\begin{enumerate}
    \item Survey of equitability metrics that are relevant to the domain of finance and equity markets, with the contribution of a new entropy metric that measures individual fairness. 
    \item Formulation of the problem of learning an equitable MM policy as a multi-objective RL problem, and proposal of an effective exploration scheme for training a MM agent.
    \item Demonstrated that the addition of an equitability reward in learning a policy for the MM helps improve the trade-off between equitability and profitability.
    \item Detailed analysis of the policy learnt using RL by varying the importance given to equitability and inventory objectives.
\end{enumerate}

\subsection{Limit Order Books}
\begin{figure}[h]
    \centering
    \includegraphics[width=\linewidth,height=1.5in]{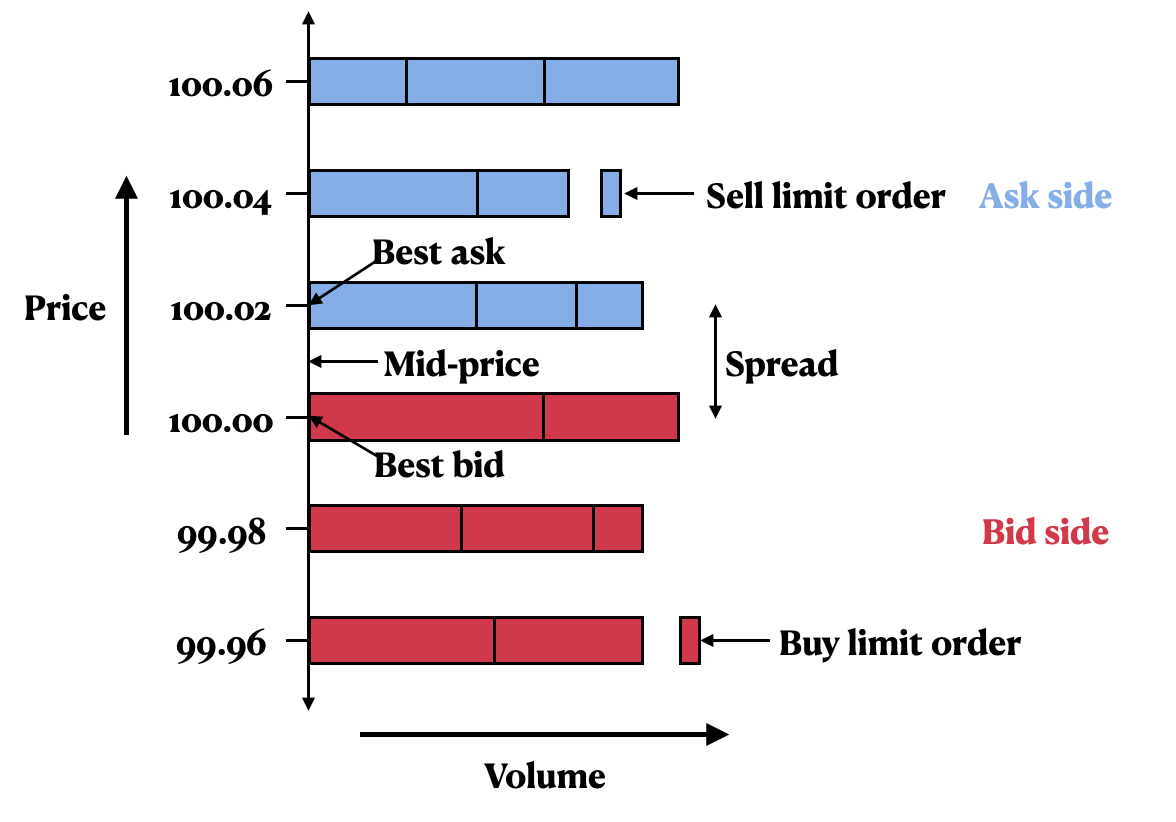}
    \caption{Snapshot of a limit order book}
    \label{fig:lob}
\end{figure}

More than half of the markets in the financial world today use a limit order book (LOB) mechanism to facilitate trade \cite{gould2013limit,Rosu_adynamic}. An LOB is the set of all orders in the market at the current time, with each order represented by a direction of trade - buy or sell (equivalently bid or ask), price and order size. There are two main order types of interest - market orders and limit orders. A \textit{market order} is a request to buy or sell instantaneously at the current market price. On the other hand, a \textit{limit order} is a request to buy or sell at a price better than or equal to its specified price. Therefore, limit orders face the risk of not being executed instantaneously and joins other limit orders in the queue at its price level. There are a fixed number of price levels in the LOB, with the gap between subsequent prices called the \textit{tick size}. The arithmetic mean of the best bid and best ask prices in the LOB is called the \textit{mid-price} for the stock. And, the difference between the best ask and best bid is called the \textit{spread}. An example snapshot of an LOB is shown in Figure \ref{fig:lob}. 

\subsection{Simulator}
In order to evaluate our MM's impact on equitability, we employ a multi-agent LOB simulator [citation redacted]. It provides a selection of background trading agent types described in section \ref{subsubsec:market_participants} and a NASDAQ-like exchange agent which lists securities for trade against an LOB with price-then-FIFO matching rules. It is also equipped with a simulation kernel that manages the flow of time and handles all inter-agent communication. Trading agents may not inspect the state of the exchange directly, but can send messages to request order book depth, obtain last trade prices, or place or cancel limit orders through the kernel, which imposes delays for computational effort and communication latency. Time proceeds in nanoseconds and all securities are priced in cents. This discrete event simulation mechanism permits rapid simulation despite the fine time resolution, as periods of inactivity can be ``skipped over'' without computational effort.

\subsubsection{Background trading agents\label{subsubsec:market_participants}}

The simulator includes agents with different trading behaviors and incentives.

{\bf Value Agents:} 
The value agents are designed to simulate fundamental traders that trade in line with their belief of the exogenous stock value (which we call {\it fundamental} price), without any view of the LOB microstructure \cite{kyle1985continuous}. The fundamental price represents the agent's understanding of the outside world (e.g. earnings reports, macroeconomic events, etc) \cite{wang2017spoofing}, \cite{wah2017welfare}. In this paper, we model the fundamental price of an asset by its historical mid-price series. Note however that significant deviations between the fundamental and the simulated mid-price are possible since agent interactions ultimately define the simulated mid-price. Each value agent arrives to the market multiple times in a trading day according to a Poisson process, and chooses to buy or sell a stock depending on whether it is cheap or expensive relative to its noisy observation of the fundamental. Upon determining the side of its order, the value agent places a limit order at a random level either inside the spread or deeper into the LOB. Therefore, value agents assist price formation in the LOB by bringing in external information. 

{\bf MM Agent:} MMs play a central role as liquidity providers by continuously quoting prices on both sides of the LOB, and earn the spread if orders execute on both sides. MMs act as intermediaries and essentially eliminate \textit{air pockets} between existing buyers and sellers. They also make markets more liquid and enable investors to trade large quantities with smaller price moves \cite{high_freq_fairness}. In this work, we define a MM by the {\it{stylized parameters}} that follow from its regulatory definition. While most realistic MM models have adverse selection and/or inventory control mechanisms that are intended to boost the MM's profitability (e.g., \cite{AvellanedaStoikov}), we however highlight that our definition is model-independent and is intended to explore the effect of its stylized properties on market equitability. Our MM definition is similar in spirit to that of \cite{wah2017welfare} and \cite{chakraborty2011market}, barring the fact that our MM does not use any reference price series to determine its mid-price. It instead adapts to the mid-price in the LOB.

{\bf Momentum Agents:} The momentum agents follow a simple momentum strategy, waking up at a fixed rate and observing the mid-price each time. They compare a long-term average of the mid-price with a short-term average. If the short-term average is higher than the long-term average, the agent buys since the price is seen to be rising. On the other hand, if the short-term average is lower than the long-term average, the agent sells since it believes that the price is falling.

{\bf Consumer Agents:} Consumer agents are designed to emulate consumer agents who trade on demand without any other considerations (\cite{kyle1985continuous}). Each consumer agent trades once a day by placing a market order of a random size in a random direction. Consumer agents arrive to the market at times that are uniformly distributed throughout the trading day. Our focus here is to numerically estimate the equitability of outcomes of consumer agents in the simulated environment described above.

\subsection{Equitability and metrics\label{subsec:metrics}}

Equitability has conventionally been studied in political philosophy and ethics \cite{johnrawls}, economics \cite{hmoulin}, and public resource distribution \cite{peytonyoung}. In recent times, there has been a renewed interest in quantifying equitability in classification tasks \cite{dwork2012fairness}. In finance, risk is synonymous with volatility since volatile markets lead to some investors doing well with others doing poorly \cite{oregonlaw}. For instance, if a consumer agent places a trade during a period of market instability when asset price moves sharply in the direction opposite to the trade and then rebounds in seconds (called mini-flash crash), the agent's execution may be perceived to be inequitable as compared to other agents who were luckier to trade in more stable markets. Accordingly, regulators are interested in reducing market volatility to facilitate equitability from the perspective of equality of market outcomes.

For this work, we are interested in the notion of individual fairness which advocates the sentiment that \textit{any two individuals who are similar with respect to a task should be treated similarly}. We now collect some metrics for individual fairness taking inspiration from those used to quantify income inequality in populations \cite{inequalityindices}, and those from information theory. The Theil index measures the \textit{distance} a population is away from the ideal egalitarian state of everyone having the same income \cite{htheil}. It belongs to the family of generalized entropy indices (GEI) that satisfy the property of subgroup-decomposability, i.e. the inequality measure over an entire population can be decomposed into the sum of a between-group unfairness component (similar to group fairness metrics in \cite{dwork2012fairness}, \cite{aif360}) and a within-group unfairness component. The Gini index is another popular metric of income inequality that measures how far a country's income distribution deviates from a totally equal distribution \cite{gini}. 

Another metric is based on information entropy, which is a measure of uncertainty of a random variable. Entropy has been commonly used as a metric for equitable resource allocation in wireless networks \cite{FairnessMetric}, a measure of income inequality and an indicator of population diversity in the social sciences \cite{segregation}, \cite{diversity}, \cite{Balch2000HierarchicSE}, \cite{TuckerThesis}. Since we have finitely many samples $y_1,\cdots,y_n$ from the distribution of individual outcomes, we estimate the entropy of individual outcomes by binning them into $K$ bins. Subsequently, we compute the entropy estimate $H_K(Y)=-\sum_{k=1}^Kp_k\log p_k$ where $p_k$ denotes the empirical frequency of outcomes in the $k^{th}$ bin. Formally, the equitability metric based on estimating information entropy from samples $y_1,\cdots,y_n$ is given by\begin{align}
    \mathcal{L}_{\mathrm{Ent}}(y_1,\cdots,y_n)=1-\frac{H_K(Y)}{\log K}=1-\frac{-\sum_{k=1}^Kp_k\log p_k}{\log K}\label{eq:ent}
\end{align}
Note that the scaling factor of $\log K$ corresponds to the entropy of a uniform distribution over the outcomes, and is most inequitable.

\subsection{Motivation\label{subsec:motivation}}

Per its regulatory definition, the MM acts as a liquidity provider by placing limit orders on both sides of the LOB with a constant arrival rate. At time $t$, it starts by cancelling any of its unexecuted orders. Let the levels of the LOB on both bid and ask sides be indexed from $0$ to $N$, with level $0$ corresponding to the innermost LOB levels. Let $b_t$ and $a_t$ denote the best bid and best ask respectively. The MM looks at the mid-price $m_t:=\frac{a_t + b_t}{2}$, and places new price quotes of constant size $K$ at $d$ price increments around $m_t$. That is, it places bids at prices $m_t-s_t-d, \ldots, m_t-s_t$ and asks at prices $m_t + s_t, \ldots, m_t + s_t +d$, where $d$ is the {\it{depth}} of placement and $s_t$ is the {\it{half-spread}} chosen by the MM at time $t$. As in NASDAQ, the difference between consecutive LOB levels is one cent. Such a stylised market maker has been shown to be profitable over a sufficiently long time horizon, when the mid-price follows an  Ornstein-Uhlenbeck Process in \cite{chakraborty2011market}.

We consider two versions of our MM agent: one that posts liquidity at a constant half-spread $s_t$ for all times $t$; and one that adapts its half-spread to its observation of the LOB at time $t$, i.e. with $s_t = \frac{a_t-b_t}{2}$. We are interested in examining the effect of the MM parameters of half-spread and depth on equitability for consumer agents. To do so, the half-spread $s_t$ is varied across values in $\lbrace1,2,\frac{a_t-b_t}{2}\rbrace$ cents, and the depth is varied across values in $\lbrace1,2\rbrace$ cents. For each (half-spread,depth) pair, we collect 20 samples of the resulting MM profit and equitability as defined by the information entropy metric in equation (\ref{eq:ent}). The green dots in figure \ref{fig:mmf01} are the average values of MM profits and equitability for the above range of spread and depth parameters. We observe a negative correlation between the two variables of interest, as represented by green line fit to the green dots. This discovery inspired us to think of a way to improve MM profits while maintaining a high level of equitability to consumer agents. We formulate this as an RL problem directed at optimizing for both MM profits and market equitability to consumer agents, using our multi agent simulator to play out interactions between traders in the market.

\section{Problem Setup\label{sec:problem}}
Consider a market configuration comprising a group of consumer agents trading alongside more intelligent investors, and a MM that provides liquidity. We are interested in using RL to improve the profits made by the MM, while guaranteeing a high degree of equitability to consumer agents. We formulate this as an RL problem by representing the market with different trading agents (including the MM) as a Markov Decision Process (MDP). An MDP is a tuple $(\mathcal{S},\mathcal{A},\mathcal{T},R,\gamma,T)$ comprising the state and action spaces along with a model of the environment (either known or unknown), the reward function. discount factor and the time horizon for the planning/control problem. The state for our MDP captured the states of both the market and the MM as
\begin{align}
\begin{bmatrix}inventory&imbalance&spread&midprice\end{bmatrix}\label{eq:state}
\end{align}
where $inventory$ is number of shares held in the MM's inventory, $imbalance=\frac{total\ bid\ volume}{total\ bid\ volume +total\ ask\ volume}$ is the volume imbalance in the order book, $spread$ is the current market spread, and $midprice$ is the current mid-price of the stock. The actions include trading actions of the MM\begin{align}
    \begin{bmatrix}s_t&d\end{bmatrix}\nonumber
\end{align}
where $s_t$ is the half-spread and $d$ is the depth of orders placed by the MM as described in section \ref{subsec:motivation}. In order to quantify the reward function, we need a way to numerically quantify equitability alongside MM profits. 

Quantifying equitability essentially involves determining if the consumer agents are treated \textit{similarly} to one another, where \textit{similarity} is ascertained with respect to the task at hand. In this paper, we propose a method for quantifying equitability in consumer agent outcomes by looking at the distribution of differences between an agent's traded price and asset price $T$ seconds afterwards. The only incentive consumer agents have to trade is demand, and outcomes should not be contingent on the time when their trades are placed. Formally, let a consumer agent $i$ execute a trade at price $p_t$ at time $t$. Define the $T$-period return $r_{i,T}$ of consumer agent $i$ for some $T \geq t$ and future price $p_T$ as\begin{align}
r_{i,T}=\begin{cases}
p_T - p_t  \textnormal{ if agent $i$ executes a buy order} \\
p_t - p_T  \textnormal{ is agent $i$ executes a sell order}
\end{cases}\label{eq:returns}
\end{align}

Any of the metrics described in \ref{subsec:metrics} could be used to quantify individual fairness based on the aforementioned price returns, but the information entropy metric (\ref{eq:ent}) was seen to be more preferable than others. This is because the GEI and Theil index are undefined (for non-even values of the GEI parameter $\alpha$) when the returns are negative, with the GEI giving spiky values hindering learning for $\alpha=2$. Since the consumer agents arrive at random times during a trading day, their returns are accumulated and the equitability reward is computed as the change in information entropy every $N>1$ time steps as
\begin{align}
&R_\mathrm{Equitability}^t\nonumber\\
&=\begin{cases}\mathcal{L}_{\mathrm{Ent}}\big(y_1,\cdots,y_{mN}\big)-\mathcal{L}_{\mathrm{Ent}}\big(y_1,\cdots,y_{(m-1)N}\big)\textnormal{ if }t=mN+1\\
    0\textnormal{ otherwise}\end{cases}\nonumber\\
&=\begin{cases}
    \frac{H_K(y_1,\cdots,y_{(m-1)N})}{\log K}-\frac{H_K(y_1,\cdots,y_{mN})}{\log K}
    \textnormal{ if }t=mN+1\\
    0\textnormal{ otherwise}
    \end{cases}\label{eq:r_fairness}  
\end{align}
where $m\in\lbrace1,2,\cdots\rbrace$ with $H(y_0):=0$. The need for using a change in the entropy metric is justified by looking at the objective of the RL problem over the entire trading day. Before doing that, we quantify the reward for profits of the MM.

The reward from profits and losses (PnL) of the MM comrpise a weighted combination of those from holding a certain inventory, and those from making the latest trade\begin{align}
    R_{\mathrm{PnL}}=\bar\eta\cdot \mathrm{inventory\ PnL}+\mathrm{matched\ PnL}\label{eq:r_pnl}
\end{align}
where inventory PnL=inventory$\times$change in mid-price and matched PnL refers to profits made by capturing the spread. The weighting factor for the inventory PnL $\bar\eta$ (called \textit{inventory-weight}) is chosen to be in $(0,1)$ due to the ensuing observation. Our MM is intended to make profits by capturing the spread upon executing orders on both sides of the LOB. But, there is always a risk of collecting inventory from one (side) of the orders not being executed, and the inventory PnL incentivizing the MM to exploit price trends. By weighing down the inventory component of PnL, we focus on learning a MM that profits from capturing the spread while also enabling a more stable learning setup as in \cite{SpoonerMarketMaking}.

Since we have two (potentially competing) objectives of profits and equitability, this becomes a multi-objective RL (MORL) problem similar to those encountered in robotics \cite{morlrobotics}, economic systems \cite{morleconomic} and natural resource allocation \cite{morlresourceallocation}. The field of MORL involves learning to act in environments with vector reward functions $\mathbf{R}:\mathcal{S}\rightarrow\mathbb{R}^n$ \footnote{The \textbf{bold face} represents a vector.}. Since the value functions in MORL induce only a partial ordering even for a given state \cite{morlsurvey}, additional information on the relative importance of the multiple objectives is given in the form of a scalarization function. To understand the relationship between profits and equitability in this paper, we found it sufficient to use a linear scalarization function. Thus, the combined reward function for our RL problem is given by\begin{align}
    R=R_{\mathrm{PnL}}+\eta\cdot R_{\mathrm{Equitability}}\label{eq:reward}
\end{align}
where $\eta\geq0$ is called the \textit{equitability-weight}, and has the interpretation of monetary benefits in \$ per unit of equitability. (\ref{eq:reward}) is also motivated from 
a constrained optimization perspective as being the objective in the unconstrained relaxation \cite{boyd2004convex} of the problem $\max R_{\mathrm{PnL}}\textnormal{ s.t. }R_\mathrm{Equitability}\geq c$. Usage of a linear scalarization function also helps us analyse the variation in equitability and profits of the learnt policy, as $\eta$ is varied. 

With the rewards defined in equations (\ref{eq:reward}), (\ref{eq:r_pnl}) and (\ref{eq:r_fairness}), discount factor $\gamma=1$ and the MM starting from the same initial state every trading day (i.e, same amount of money and shares), the RL objective under a policy $\pi$ evaluates to 
\begin{equation}
    \begin{aligned}
        \bar\eta\cdot\mathbb{E}\left[\sum_{t=0}^{T-1}\mathrm{inventory\ PnL}\right]+&\eta\cdot\mathbb{E}\left[\frac{-H_K\left(y_1,\cdots,y_{L-2}\right)}{\log K}\right]\\
        +&\mathbb{E}\left[\sum_{t=0}^{T-1}\mathrm{matched\ PnL}\right]
    \end{aligned}\label{eq:weighted_obj}
\end{equation}

\textbf{Hence, the objective of the equitable MM is to learn a trading policy that maximizes its PnL over a given time horizon, while minimizing the entropy of consumer agent returns.}

Since it is straightforward to compute the optimal policy given the state-action value function or Q function (as the argument that maximizes the Q value in each state), we use the Q learning algorithm that iteratively updates estimates of the optimal Q function \cite{watkins1992q}. The Q function is defined under a policy $\pi$ as the long-term value of taking action $a$ in state $s$ and following policy $\pi$ thereafter \begin{equation}
    \begin{aligned}
        Q^\pi(s,a)=\mathbb{E}\bigg[\sum_{t=0}^{T-1}
\gamma^tR(s_{t+1})&\bigg|s_{t+1}\sim\mathcal{T}\big(\cdot|s_t,a_t\big),a_t\sim\pi(\cdot|s_t),\\
&s_0=s,a_0=a\bigg]
    \end{aligned}\label{eq:q}
\end{equation}
Since we consider a finite set of (discretized) states and actions, the Q function is a table of values for all possible (state,action) pairs. A useful property of learning such a tabular Q function is that the resulting policy is more easily interpretable as opposed to using function approximators such as neural networks. 

The Q learning algorithm uses a stochastic approximation method to estimate the optimal Q function from the Bellman equation \cite{robbins1951stochastic,gladyshev1965stochastic}. It updates the Q values in the $n^{th}$ episode as \begin{equation}
\begin{aligned}
    &Q_n(s,a)\\
&=\begin{cases}
\big(1-\alpha_n\big)Q_{n-1}(s,a)+\alpha_n\big(R(s')+\gamma V_{n-1}(s')\big)\textnormal{ if }(s_n,a_n)=(s,a)\\
Q_{n-1}(s,a)\textnormal{ otherwise}
\end{cases}
    \end{aligned}\label{eq:qlearning}
\end{equation}
where $s'$ is the next state when action $a$ is taken in state $s$, $\alpha_n$ is the learning rate in the $n^{th}$ episode, and $V_{n-1}(s')=\max_{a'}Q_{n-1}(s',a')$. The iterates $Q_n(\cdot,\cdot)$ are guaranteed to converge to the optimal Q values if the rewards are bounded, the learning rates satisfy $0\leq\alpha_n<1$, and the sequence of learning rates $\lbrace\alpha_n\rbrace$ satisfy $\sum_n\alpha_n=\infty$ and $\sum_n\alpha_n^2<\infty$. The optimal policy is then given by $\pi^\star(s)=\arg\max_aQ^\star(s,a)\ \forall s\in\mathcal{S}$.

\section{Experiments}
\subsection{Non learning experiments}

\subsubsection{Relationship between profits and equitability for different consumer agent order sizes}
We saw a negative correlation between MM profits and market equitability upon varying the MM half spread and depth in Figure \ref{fig:mmf01}. Another important parameter is the order size of consumer agents, since equitability for bigger (higher order volume) players potentially differs from that for smaller players, even though they are of the same trading type. Therefore, we now vary the parameters of half-spread and depth of MM orders, as well as consumer agent order size in order to examine their impact on market equitability to consumer agents. The half-spread takes values in $\lbrace2,5,10,20,\frac{a_t-b_t}{2}\rbrace$ cents, the depth takes values in $\lbrace1,2,5,10,15\rbrace$ cents, and consumer agent order size takes values in $\lbrace5,10,30,50,100\rbrace$. For each (half-spread,depth,order size) triple, we collect 200 samples of the resulting MM profit and equitability as defined by the information entropy metric in equation (\ref{eq:ent}). 

\begin{figure}[h]
    \centering
    \includegraphics[width=.8\linewidth,height=2in]{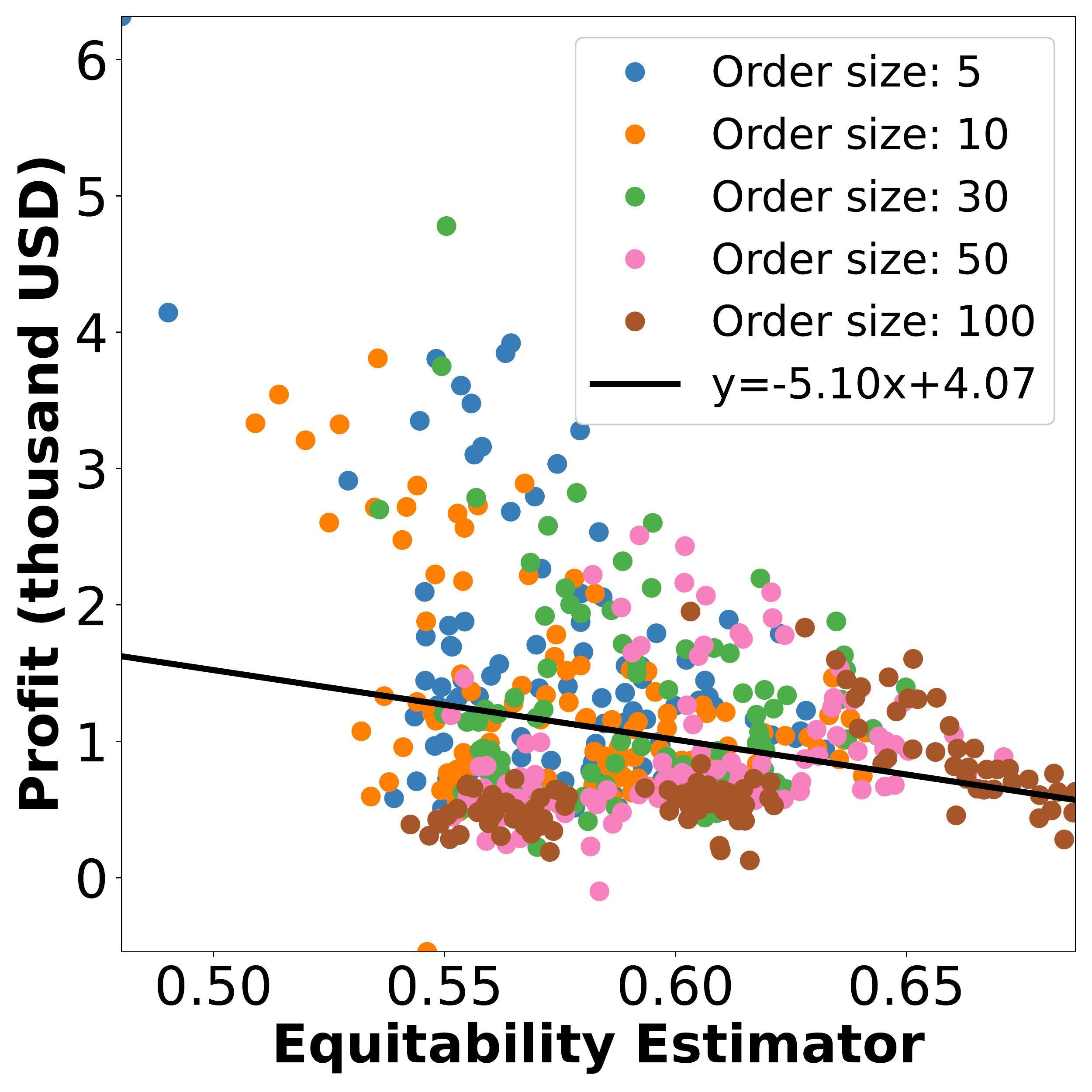}
    \caption{Profits versus equitability for all order sizes}
    \label{fig:mmf1}
    \includegraphics[height=3.5in,width=.8\linewidth]{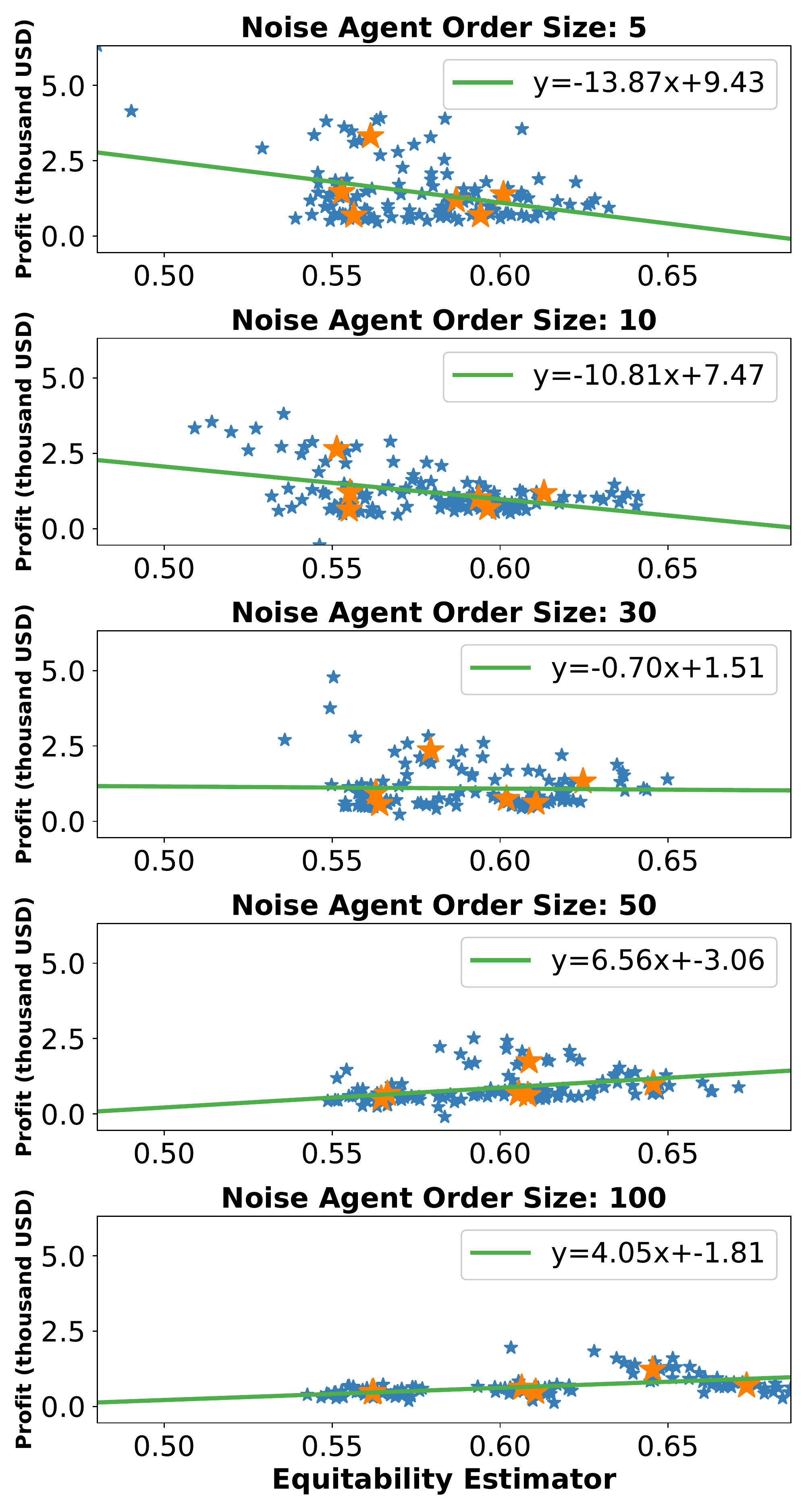}
    \caption{Profits versus equitability for each order size}
    \label{fig:mmf2}
\end{figure}

Figure \ref{fig:mmf1} is a plot of MM profits versus equitability for all order sizes along with a common (black) regression line. On examining the scattered points, we see that equitability increases with order size. Figure \ref{fig:mmf2} is a plot of profits versus equitability with subplots representing each order size with regression lines for their corresponding order sizes. The larger orange stars represent the average MM profit and equitability for a fixed (half-spread,depth). We see an improvement in the correlation between profits and equitability with increase in order size. This confirms our intuition that the stylized MM is more equitable to consumer agents that place large orders than those that place smaller ones.

\subsubsection{Relationship between profits and equitability for different sources of liquidity}

\begin{figure}[h]
    \centering
    \includegraphics[width=.8\linewidth,height=3.7in]{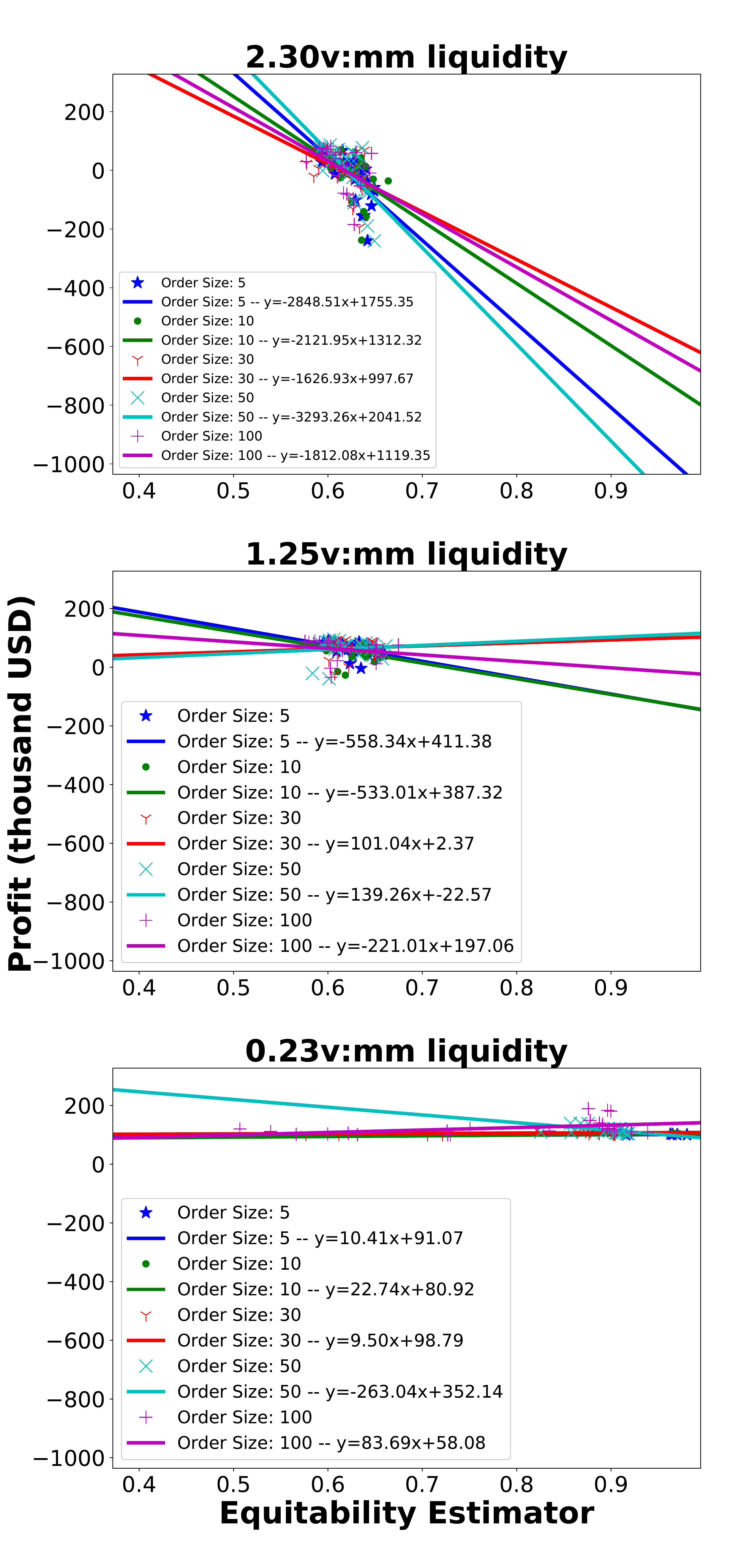}
    \caption{Varying liquidity from value agents}
    \label{fig:mmf3}
\end{figure}

We note that the consumer agents we consider in this paper are meant to refer to individual traders that place orders purely based on demand. They do not use external information or any intelligence in trading, unlike value agents and momentum agents described previously. Once the value agents bring in fundamental information into the market through their trading strategies, other background agents can observe their actions to infer the fundamental value. Thus, intelligent traders cause information flow in the market from their access to external information. We now look at the influence of external information on market equitability by varying the amount of liquidity provided by value agents as a fraction of that provided by our stylized MM. Figure \ref{fig:mmf3} is a plot of MM profits versus equitability to consumer agents for decreasing amounts of liquidity provided by value agents as a fraction of that provided by our MM. Each sub-figure also has regression lines corresponding to different order sizes for consumer agents. We see that the correlation improves with decrease in value agent liquidity for each consumer agent order size, i.e. a market with more liquidity coming from our stylized MM is more equitable to consumer agents, as opposed to one where more liquidity comes from value agents. Hence, it is more equitable to have less proprietary information flowing into the market since the consumer agents (from the way we have defined them) don't use that information themselves. 

\subsection{Learning experiments}
We first discretize our states and actions so as to make the problem suited to using RL methods for finite MDPs. An important requirement for the functioning of the tabular Q learning algorithm is that there be enough visitations of each (state,action) pair. Accordingly, we pick our state and action discretization bins by observing the range of values taken in a sample experiment. The discrete state computed after measuring the state in (\ref{eq:state}) is $s=\begin{bmatrix}
    s_{\mathrm{inventory}}&s_{\mathrm{imbalance}}&s_{\mathrm{spread}}&s_{\mathrm{midprice}}
    \end{bmatrix}$ where \begin{align*}
s_{\mathrm{inventory}}&=\begin{cases}0\textnormal{ if }\mathrm{inventory}<-10I\\1\textnormal{ if }-10I\leq\mathrm{inventory}<-5I\\2\textnormal{ if }-5I\leq\mathrm{inventory}<0\\3\textnormal{ if }0\leq\mathrm{inventory}<5I\\4\textnormal{ if }5I\leq\mathrm{inventory}<10I\\5\textnormal{ if }\mathrm{inventory}\geq10I\end{cases}\\
    s_{\mathrm{imbalance}}&=\begin{cases}0\textnormal{ if }0\leq\mathrm{imbalance}<0.25\\1\textnormal{ if }0.25\leq\mathrm{imbalance}<0.5\\2\textnormal{ if }0.5\leq\mathrm{imbalance}<0.75\\3\textnormal{ if }0.75\leq\mathrm{imbalance}\leq1\end{cases}\\
    s_{\mathrm{spread}}&=\begin{cases}0\ \mathrm{if}\ 0\leq\mathrm{spread}<2\\1\ \mathrm{otherwise}\end{cases}\\
    s_{\mathrm{midprice}}&=\begin{cases}0\textnormal{ if }\mathrm{midprice}<m_0\\1\textnormal{ if }\mathrm{midprice}\geq m_0\end{cases}
\end{align*}
where $I$ and $m_0$ are constants. The discrete actions taken by the MM are of the form $a=\begin{bmatrix}a_\mathrm{mm-spread}&a_\mathrm{mm-depth}\end{bmatrix}$ to encode the spread and depth of orders placed by the MM as\begin{align*}
    a_{\mathrm{mm-spread}}&=\begin{cases}
    0\leftrightarrow\mathrm{mm-spread}=\mathrm{current\ spread}\\
    1\leftrightarrow\mathrm{mm-spread}=1\\
    2\leftrightarrow\mathrm{mm-spread}=2
    \end{cases}\\
    a_{\mathrm{mm-depth}}&=\begin{cases}
    0\leftrightarrow\mathrm{mm-depth}=1\\
    1\leftrightarrow\mathrm{mm-depth}=2
    \end{cases}
\end{align*}
The discretization bins for consumer agent returns are chosen based on their observed values in a sample experiment with $K=12$ as $(-\infty,-10^{5})\cup[-10^5,-10^4)\cdots\cup[10^4,10^5)\cup[10^5,\infty)$ with intervals corresponding to bins for computing empirical probabilities. 
\subsubsection{Training and convergence}
The Q learning algorithm (\ref{eq:qlearning}) is used to compute estimates of the optimal Q function for the MDP described above. In order to balance exploration and exploitation, we use an $\epsilon$ - greedy approach to choose the next action in a given state. Hence, in episode $n$, the next action is chosen to be that which maximizes the current Q function estimate $Q_n$ with a probability of $1-\epsilon_n$, and is chosen randomly with probability $\epsilon_n$. We call $\epsilon_n$ the exploration parameter that is decayed as episodes progress. An important assumption underlying the convergence results for the tabular Q learning algorithm is that we have \textit{adequate} visitation of each (state, action) pair. We found that having a dedicated exploration phase where the exploration parameter $\epsilon$ and learning rate $\alpha$ are kept constant at high values helps us in obtaining these visitations. 

Training is divided into three phases - pure exploration, pure exploitation and convergence phases. During the pure exploration phase, $\alpha_n$ and $\epsilon_n$ are both held constant at high values in order to facilitate the visitation of as many state-action discretization bins as possible. During the pure exploitation phase, $\epsilon_n$ is decayed to an intermediate value while $\alpha_n$ is held constant at its exploration value so that the Q Table is updated to reflect the one step optimal actions. After the pure exploration and pure exploitation phases, we have the learning phase where both $\alpha_n$ and $\epsilon_n$ are decayed to facilitate convergence of the Q learning algorithm. Note that each training episode corresponds to one trading day from 9:30am until 4:30pm. Since the MM wakes up every five seconds, this gives $T=5040$ steps per episode. The precise numerics of our learning experiments are given in Table \ref{tab:expt_numerics}. 

\begin{table}
    \centering
    \begin{tabular}{|p{0.6\linewidth}|p{0.35\linewidth}|}\hline
        Number of pure exploration episodes & 400 \\\hline
        Number of pure exploitation episodes & 200 \\\hline
        Number of convergence episodes & 400 \\\hline
        Total number of training episodes & 1000\\\hline
        $\gamma$ & $1.0$\\\hline
        $T$ & $5040$\\\hline
        $\alpha_0=\alpha_1=\cdots=\alpha_{399}=\cdots=\alpha_{599}$  & $0.9$\\\hline
        $\alpha_{999}$ & $10^{-5}$\\\hline
        $\epsilon_0=\epsilon_1=\cdots=\epsilon_{399}$ & $0.9$\\\hline
        $\epsilon_{599}$ & $0.1$ \\\hline
        $\epsilon_{999}$ & $10^{-5}$\\\hline
        $\eta$ & $\lbrace0,0.05,0.1,0.15,0.2,0.3\rbrace$\\\hline
        $\bar\eta$ & $\lbrace0,3,6,10,20,50\rbrace$\\\hline
        $I$ & 500\\\hline
        $m_0$ & $10^5$\\\hline
        $K$ & 12\\\hline
    \end{tabular}
    \caption{Numerics for experiments}
    \label{tab:expt_numerics}
\end{table}

\begin{figure}
\centering
\includegraphics[width=\linewidth,height=1.5in]{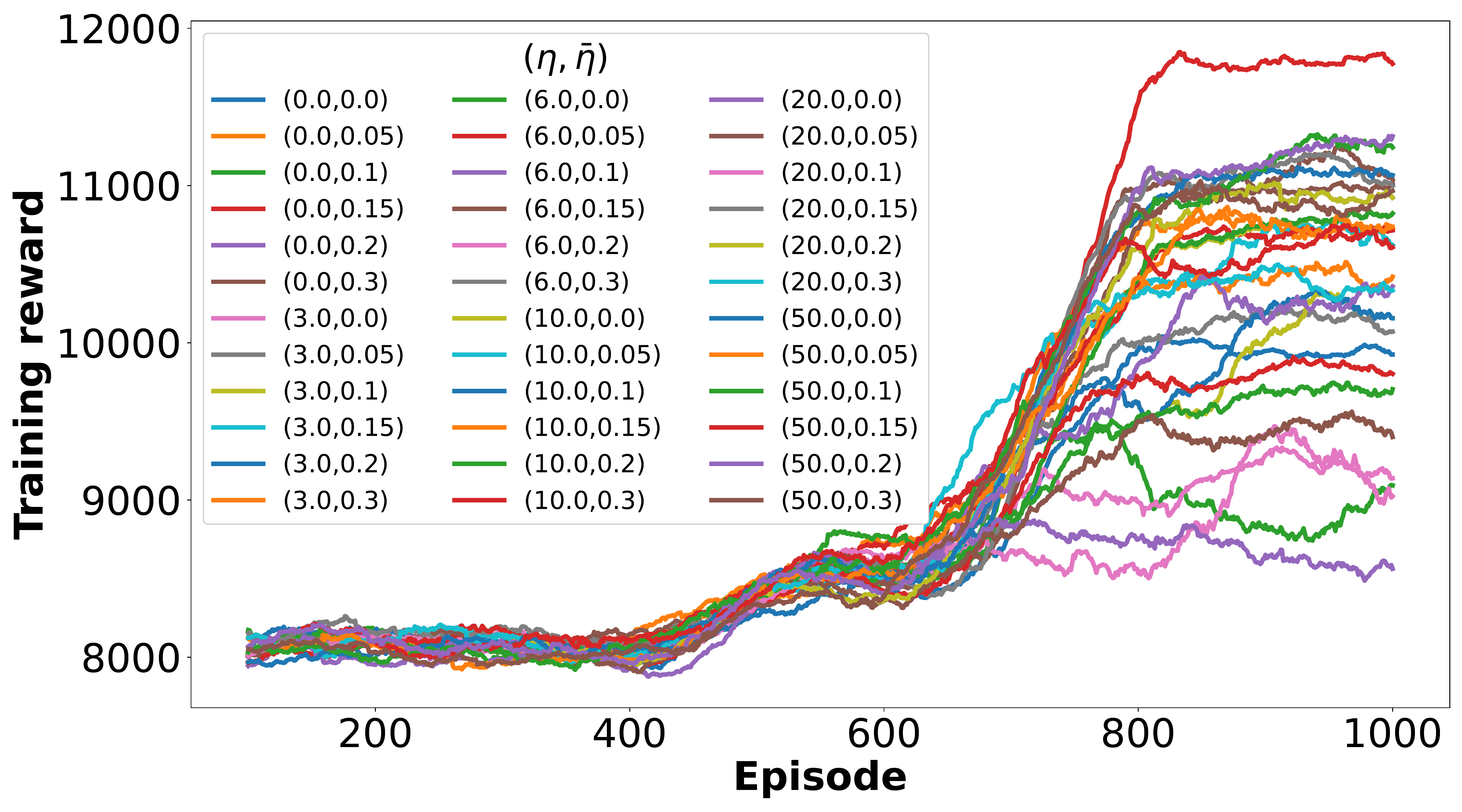}
\caption{Training rewards for $(\eta,\bar\eta)$}
\label{fig:training_rewards}    
\end{figure}

We vary the equitability weight $\eta$ and the inventory weight $\bar\eta$ over the range of values given in Table \ref{tab:expt_numerics}. The training rewards for all $(\eta,\bar\eta)$ are plotted as a function of training episodes in Figure \ref{fig:training_rewards}. 
We are able to achieve convergence for the tabular Q learning algorithm to estimate optimal MM actions in our multi-agent simulator for the range of weights $(\eta,\bar\eta)$ considered. 

\subsubsection{Understanding learnt policies for different $(\eta,\bar\eta)$}

\begin{figure}
    \centering
    \includegraphics[width=.6\linewidth,height=1in]{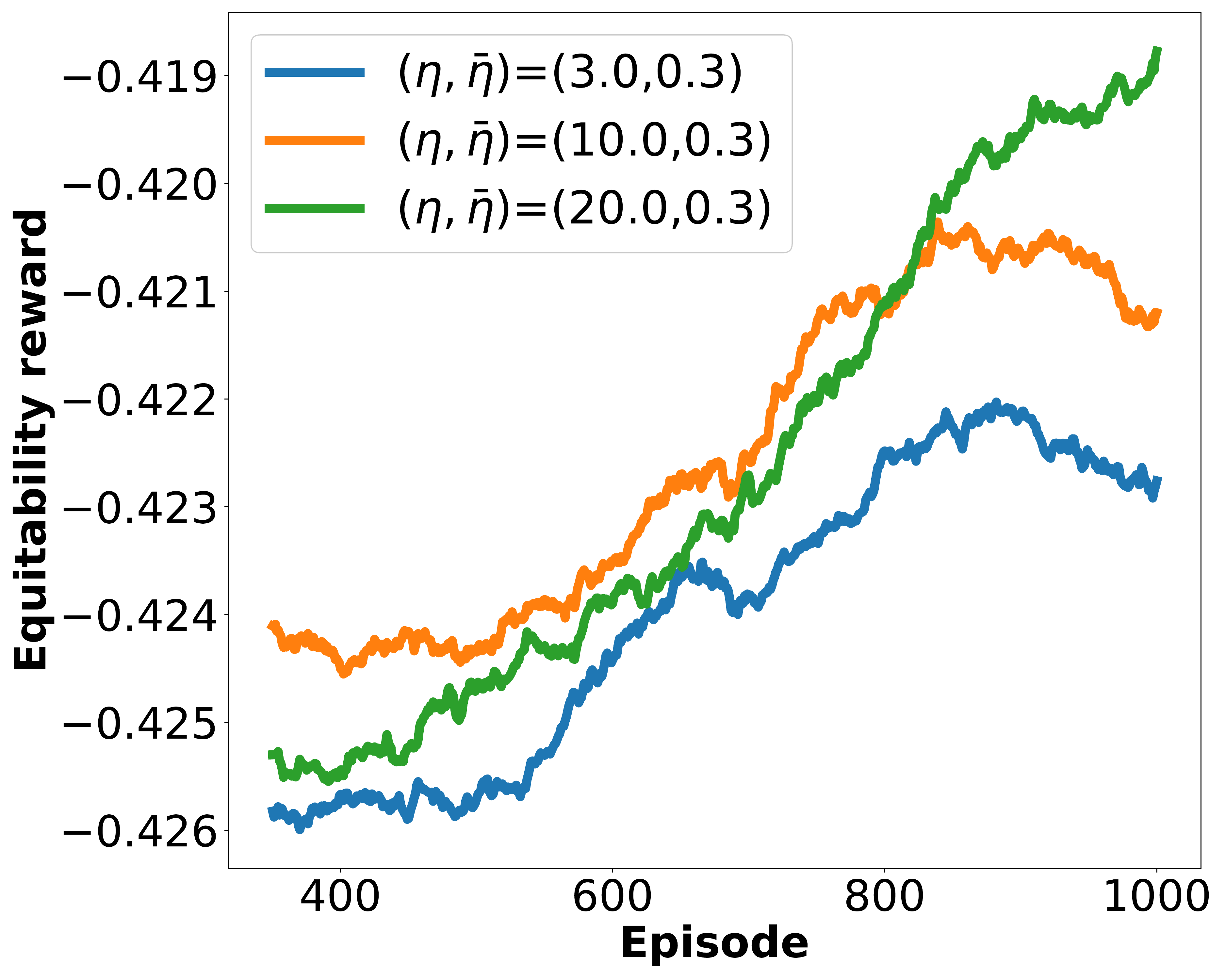}
    \caption{Equitability reward for different $\eta$}
    \label{fig:r_fair}
\end{figure}

We now attempt intuitively explaining the effect of varying the weights $(\eta,\bar\eta)$ on the policy learnt using the objective (\ref{eq:weighted_obj}). Figure \ref{fig:r_fair} shows the cumulative equitability reward $\mathbb{E}\left[\frac{-H_K\left(y_1,\cdots,y_{L-2}\right)}{\log K}\right]$ under the learnt policy for $\eta\in\lbrace3,10,20\rbrace$, and $\bar\eta=0.3$. As expected, we see that the equitability reward for the learnt policy increases as a function of the equitability weight $\eta$ (for fixed $\bar\eta$). 

\begin{figure}
    \centering
    \includegraphics[width=.9\linewidth,height=1.2in]{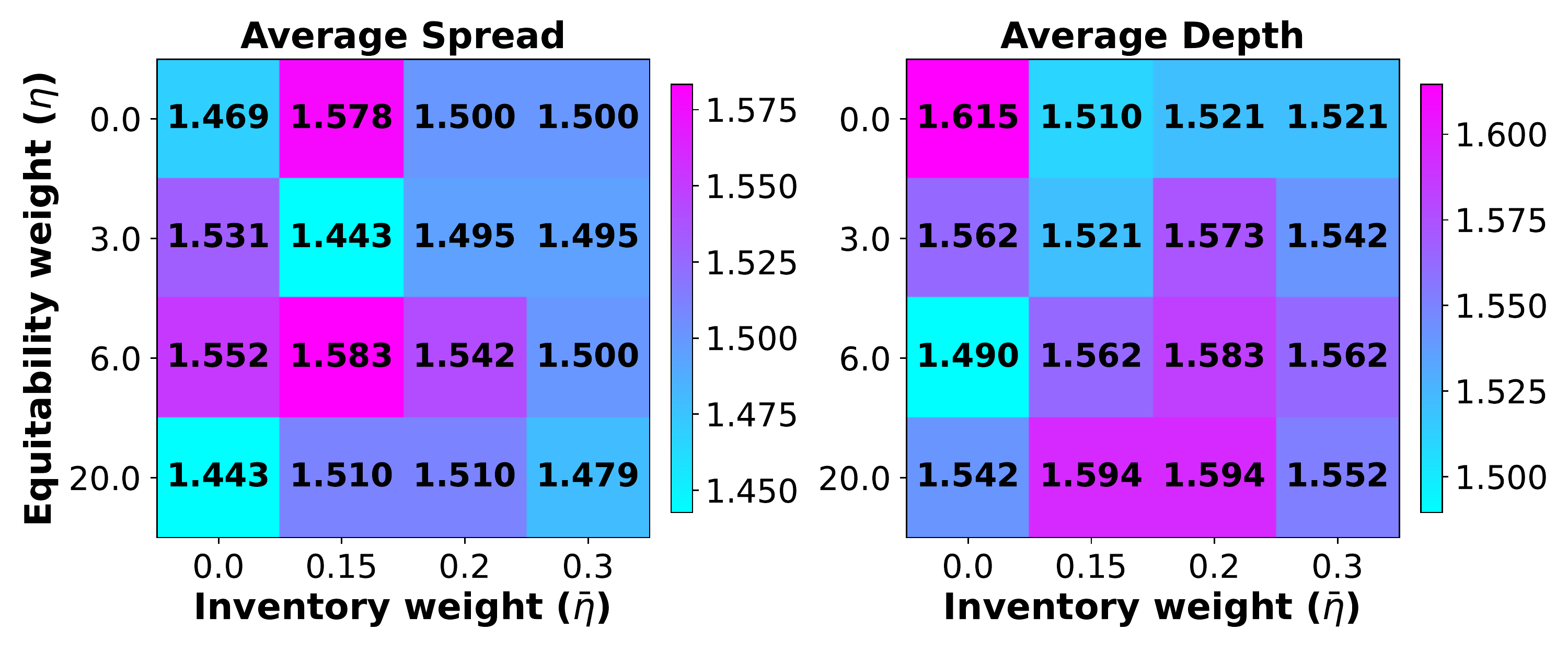}
    \caption{Average spreads and depths of orders placed by MM for different $(\eta,\bar\eta)$}
    \label{fig:avg_s_d}
\end{figure}

Recall that the learnt policy is a map from the current state to the (estimated) optimal action to be taken in that state. In our case, the learnt policy specifies the spread and depth of orders to be placed by our stylized MM based on current market conditions described by the state. We now look at the average spread and depth of orders placed by the learnt MM policy (averaged using a uniform distribution on the states) as functions of the weights $(\eta,\bar\eta)$. Figure \ref{fig:avg_s_d} shows heat maps of the average spread and depth for various values of $\eta$ and $\bar\eta$. We see that as we increase $\eta$, the MM learns to place orders at smaller average spread, and larger depths resulting in more liquid markets. And, this confirms our intuition that more liquid markets are more equitable.

\subsubsection{Effect of $(\eta,\bar\eta)$ on volatility of returns and PnL}

\begin{figure}
    \centering
    \includegraphics[width=\linewidth,height=1.2in]{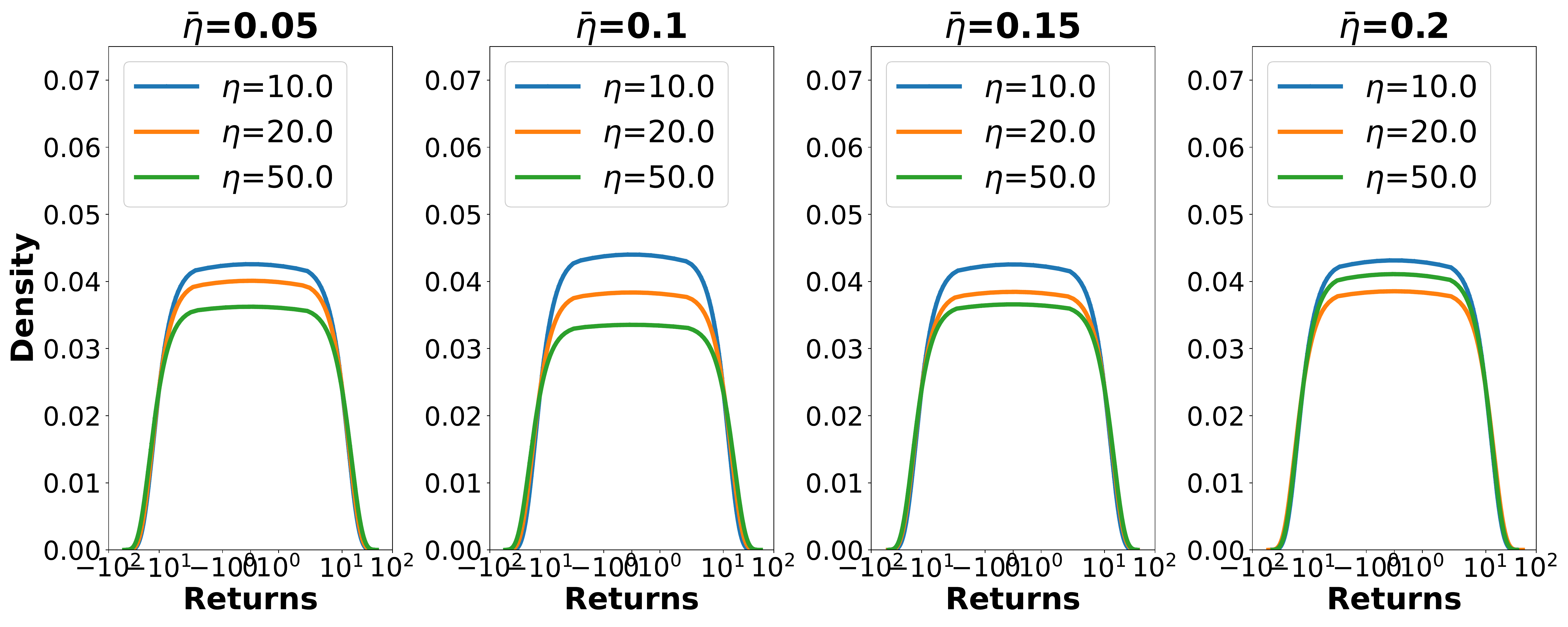}
    \caption{Distribution of consumer agent returns}
    \label{fig:returns_dist}
\end{figure}

To further explore the effects of the weights $(\eta,\bar\eta)$ on the learnt policy, we examine the price returns defined in (\ref{eq:returns}) and the cumulative MM PnL $\mathbb{E}\left[\sum_{t=0}^{T-1}\left(\mathrm{inventory\ PnL}+\mathrm{matched\ PnL}\right)\right]$. Figure \ref{fig:returns_dist} is a histogram of the resulting price returns from using the learnt policy for different $(\eta,\bar\eta)$. Each sub-figure corresponds to the distribution of returns for a fixed $\bar\eta$. Within each sub-figure, we see that the the distribution corresponding to the highest equitability weight $\eta=50$ has the lowest variance. Thus, the return volatility is seen to reduce with an increase in $\eta$, even though we do not explicitly optimize for volatility in our objective.

\begin{figure}
    \centering
    \includegraphics[width=\linewidth,height=1.2in]{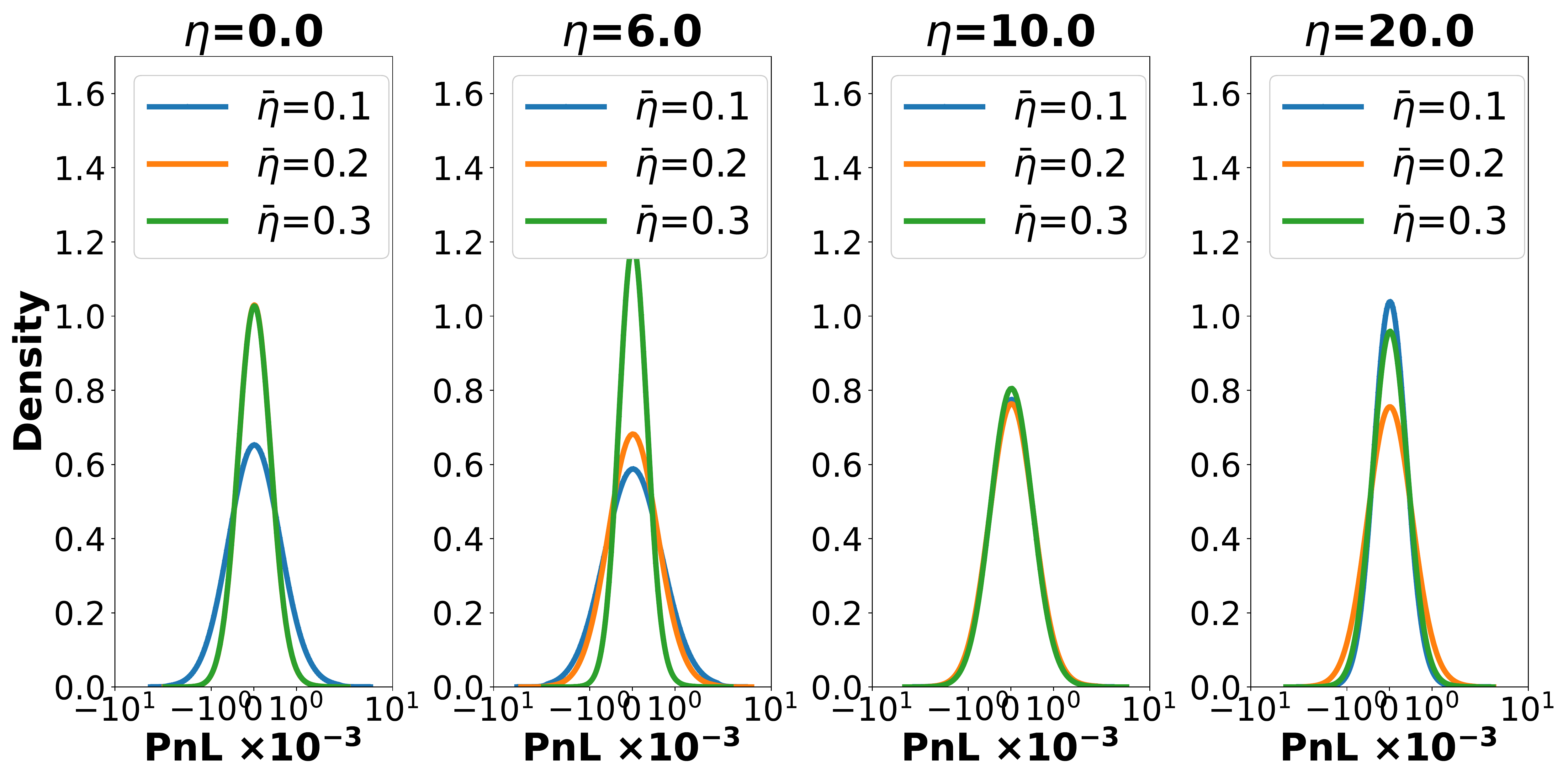}
    \caption{Distribution of MM PnL}
    \label{fig:pnl_dist}
\end{figure}

Likewise, we plot a histogram of the cumulative MM PnL that results from using the learnt policy with different $(\eta,\bar\eta)$ in Figure \ref{fig:pnl_dist}. Each sub-figure corresponds to the distribution of the PnL for a fixed $\eta$. Within each sub-figure, we see that the the distribution corresponding to the highest inventory weight $\bar\eta=0.3$ has the lowest variance. Therefore,the PnL volatility is seen to reduce with an increase in $\bar\eta$, suggesting that a high $\bar\eta$ results in a more risk sensitive MM.

\subsubsection{Effect of competing MM on learnt policy}
\begin{figure}
    \centering
    \includegraphics[width=.8\linewidth,height=1in]{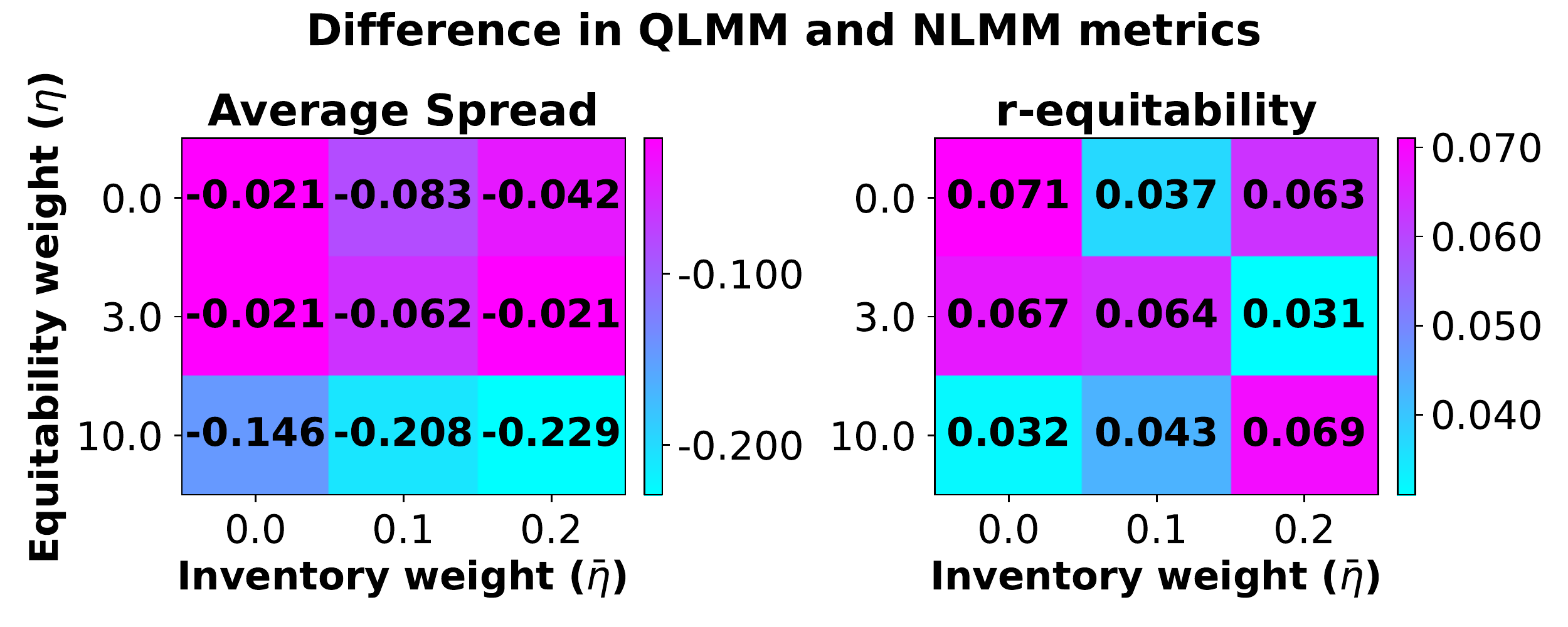}
    \caption{Effect of competing MM on learnt policy}
    \label{fig:qmm_d2}
\end{figure}

Having understood the policy learnt by our stylized MM when acting solo in the market, a natural next step would be to add a competing MM into the market and contrast the new learnt policy to the previous one. The competing MM is a non-learning based stylized MM (denoted by NLMM for non-learning MM) that posts liquidity on both sides of the spread at a fixed number of levels. Note that both MMs are made to post liquidity at the same frequency and of equal order volumes into our market. The training steps described previously are repeated for the learning based MM (denoted by QLMM for Q Learning MM), and the learnt policy is compared to that of NLMM. Figure \ref{fig:qmm_d2} is a plot of the difference in the average spread of orders and the equitability metric between QLMM and NLMM, for different $(\eta,\bar\eta)$. We see that QLMM learns to place at smaller spreads than NLMM in order to achieve a competitive advantage. We also see that QLMM is more equitable as per our entropy based equitability metric than the competing NLMM. 

\begin{figure}
    \centering
    \includegraphics[width=.8\linewidth,height=1.5in]{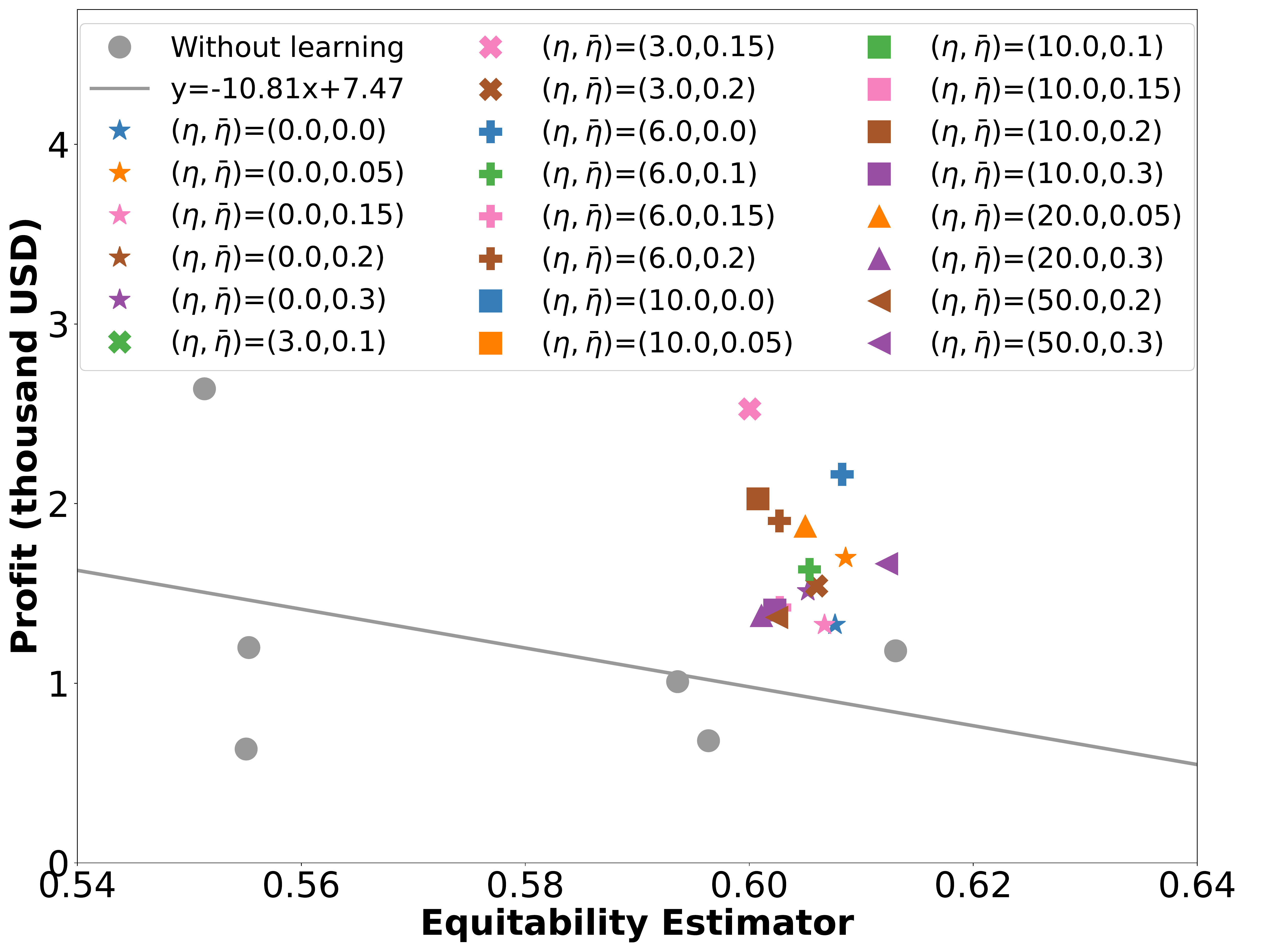}
    \caption{MM profits versus equitability to consumer agents}
    \label{fig:mmf01}
\end{figure}

The last experiment is to check if learning helped solved the problem that motivated our work in section \ref{subsec:motivation}. In order to understand the advantages of using learning to dynamically pick the spread and depth of MM orders based on current market conditions, we augment the motivating observations with the resulting MM profit and equitability to consumer agents for the learnt policies as in Figure \ref{fig:mmf01}. We see that learning helps improve the profits of the MM alongside preserving market equitability. 

\section{Conclusion and remarks}

{
In this paper, we analyzed the connection between a MM’s profitability and the equitability of outcomes and conclude
that market configurations that are more equitable are less
profitable for the MM. Not only do MMs have the capacity to strongly affect market equitability, they also lose money
for enabling it. Accordingly, regulators and exchanges may
consider incentives for MM behavior to
encourage less volatile market outcomes that are more equitable.
Note that our findings do not rely on a specific MM
model, and only use stylized properties from the regulatory
definition of a MM. We further demonstrated the ability to derive MM policies using RL that improve upon equitability outcomes of consumer agents in a simulated market. This opens up the possibility of improving on MM profits by using equitability as part of the MM's objectives.}

\begin{acks}
This paper was prepared for informational purposes by the Artificial Intelligence Research group of JPMorgan Chase \& Co\. and its affiliates (``JP Morgan''), and is not a product of the Research Department of JP Morgan. JP Morgan makes no representation and warranty whatsoever and disclaims all liability, for the completeness, accuracy or reliability of the information contained herein. This document is not intended as investment research or investment advice, or a recommendation, offer or solicitation for the purchase or sale of any security, financial instrument, financial product or service, or to be used in any way for evaluating the merits of participating in any transaction, and shall not constitute a solicitation under any jurisdiction or to any person, if such solicitation under such jurisdiction or to such person would be unlawful.
\end{acks}

\bibliographystyle{ACM-Reference-Format}
\bibliography{refs}


\begin{thebibliography}{38}


\ifx \showCODEN    \undefined \def \showCODEN     #1{\unskip}     \fi
\ifx \showDOI      \undefined \def \showDOI       #1{#1}\fi
\ifx \showISBNx    \undefined \def \showISBNx     #1{\unskip}     \fi
\ifx \showISBNxiii \undefined \def \showISBNxiii  #1{\unskip}     \fi
\ifx \showISSN     \undefined \def \showISSN      #1{\unskip}     \fi
\ifx \showLCCN     \undefined \def \showLCCN      #1{\unskip}     \fi
\ifx \shownote     \undefined \def \shownote      #1{#1}          \fi
\ifx \showarticletitle \undefined \def \showarticletitle #1{#1}   \fi
\ifx \showURL      \undefined \def \showURL       {\relax}        \fi
\providecommand\bibfield[2]{#2}
\providecommand\bibinfo[2]{#2}
\providecommand\natexlab[1]{#1}
\providecommand\showeprint[2][]{arXiv:#2}

\bibitem[\protect\citeauthoryear{Angel and McCabe}{Angel and McCabe}{2013}]%
        {high_freq_fairness}
\bibfield{author}{\bibinfo{person}{James Angel} {and} \bibinfo{person}{Douglas
  McCabe}.} \bibinfo{year}{2013}\natexlab{}.
\newblock \showarticletitle{{Fairness in Financial Markets: The Case of High
  Frequency Trading}}.
\newblock \bibinfo{journal}{\emph{Journal of Business Ethics}}
  (\bibinfo{year}{2013}).
\newblock


\bibitem[\protect\citeauthoryear{Avellaneda and Stoikov}{Avellaneda and
  Stoikov}{2008}]%
        {AvellanedaStoikov}
\bibfield{author}{\bibinfo{person}{Marco Avellaneda} {and}
  \bibinfo{person}{Sasha Stoikov}.} \bibinfo{year}{2008}\natexlab{}.
\newblock \showarticletitle{High-frequency trading in a limit order book}.
\newblock


\bibitem[\protect\citeauthoryear{Balch}{Balch}{1998}]%
        {TuckerThesis}
\bibfield{author}{\bibinfo{person}{Tucker Balch}.}
  \bibinfo{year}{1998}\natexlab{}.
\newblock \emph{\bibinfo{title}{Behavioral diversity in learning robot teams}}.
\newblock \bibinfo{thesistype}{Ph.D. Dissertation}. \bibinfo{school}{Georgia
  Institute of Technology}.
\newblock


\bibitem[\protect\citeauthoryear{Balch}{Balch}{2000}]%
        {Balch2000HierarchicSE}
\bibfield{author}{\bibinfo{person}{Tucker~R. Balch}.}
  \bibinfo{year}{2000}\natexlab{}.
\newblock \showarticletitle{Hierarchic Social Entropy: An Information Theoretic
  Measure of Robot Group Diversity}.
\newblock \bibinfo{journal}{\emph{Autonomous Robots}}  \bibinfo{volume}{8}
  (\bibinfo{year}{2000}), \bibinfo{pages}{209--238}.
\newblock


\bibitem[\protect\citeauthoryear{Bellamy, Dey, Hind, Hoffman, Houde, Kannan,
  Lohia, Martino, Mehta, Mojsilovic, et~al\mbox{.}}{Bellamy
  et~al\mbox{.}}{2018}]%
        {aif360}
\bibfield{author}{\bibinfo{person}{Rachel~KE Bellamy}, \bibinfo{person}{Kuntal
  Dey}, \bibinfo{person}{Michael Hind}, \bibinfo{person}{Samuel~C Hoffman},
  \bibinfo{person}{Stephanie Houde}, \bibinfo{person}{Kalapriya Kannan},
  \bibinfo{person}{Pranay Lohia}, \bibinfo{person}{Jacquelyn Martino},
  \bibinfo{person}{Sameep Mehta}, \bibinfo{person}{Aleksandra Mojsilovic},
  {et~al\mbox{.}}} \bibinfo{year}{2018}\natexlab{}.
\newblock \showarticletitle{AI Fairness 360: An extensible toolkit for
  detecting, understanding, and mitigating unwanted algorithmic bias}.
\newblock \bibinfo{journal}{\emph{arXiv preprint arXiv:1810.01943}}
  (\bibinfo{year}{2018}).
\newblock


\bibitem[\protect\citeauthoryear{Bone and Dragićević}{Bone and
  Dragićević}{2009}]%
        {morlresourceallocation}
\bibfield{author}{\bibinfo{person}{Christopher Bone} {and}
  \bibinfo{person}{Suzana Dragićević}.} \bibinfo{year}{2009}\natexlab{}.
\newblock \showarticletitle{GIS and Intelligent Agents for Multiobjective
  Natural Resource Allocation: A Reinforcement Learning Approach}.
\newblock \bibinfo{journal}{\emph{Transactions in GIS}} \bibinfo{volume}{13},
  \bibinfo{number}{3} (\bibinfo{year}{2009}), \bibinfo{pages}{253--272}.
\newblock
\urldef\tempurl%
\url{https://doi.org/10.1111/j.1467-9671.2009.01151.x}
\showDOI{\tempurl}
\showeprint{https://onlinelibrary.wiley.com/doi/pdf/10.1111/j.1467-9671.2009.01151.x}


\bibitem[\protect\citeauthoryear{Boyd, Boyd, and Vandenberghe}{Boyd
  et~al\mbox{.}}{2004}]%
        {boyd2004convex}
\bibfield{author}{\bibinfo{person}{Stephen Boyd}, \bibinfo{person}{Stephen~P
  Boyd}, {and} \bibinfo{person}{Lieven Vandenberghe}.}
  \bibinfo{year}{2004}\natexlab{}.
\newblock \bibinfo{booktitle}{\emph{Convex optimization}}.
\newblock \bibinfo{publisher}{Cambridge university press}.
\newblock


\bibitem[\protect\citeauthoryear{Chakraborty and Kearns}{Chakraborty and
  Kearns}{2011}]%
        {chakraborty2011market}
\bibfield{author}{\bibinfo{person}{Tanmoy Chakraborty} {and}
  \bibinfo{person}{Michael Kearns}.} \bibinfo{year}{2011}\natexlab{}.
\newblock \showarticletitle{Market making and mean reversion}. In
  \bibinfo{booktitle}{\emph{Proceedings of the 12th ACM conference on
  Electronic commerce}}. ACM, \bibinfo{pages}{307--314}.
\newblock


\bibitem[\protect\citeauthoryear{Chan and Shelton}{Chan and Shelton}{2001}]%
        {chan2001electronic}
\bibfield{author}{\bibinfo{person}{Nicholas~Tung Chan} {and}
  \bibinfo{person}{Christian Shelton}.} \bibinfo{year}{2001}\natexlab{}.
\newblock \showarticletitle{An electronic market-maker}.
\newblock  (\bibinfo{year}{2001}).
\newblock


\bibitem[\protect\citeauthoryear{Dwork, Hardt, Pitassi, Reingold, and
  Zemel}{Dwork et~al\mbox{.}}{2012}]%
        {dwork2012fairness}
\bibfield{author}{\bibinfo{person}{Cynthia Dwork}, \bibinfo{person}{Moritz
  Hardt}, \bibinfo{person}{Toniann Pitassi}, \bibinfo{person}{Omer Reingold},
  {and} \bibinfo{person}{Richard Zemel}.} \bibinfo{year}{2012}\natexlab{}.
\newblock \showarticletitle{Fairness through awareness}. In
  \bibinfo{booktitle}{\emph{Proceedings of the 3rd innovations in theoretical
  computer science conference}}. \bibinfo{pages}{214--226}.
\newblock


\bibitem[\protect\citeauthoryear{Foucault, Kadan, and Kandel}{Foucault
  et~al\mbox{.}}{2009}]%
        {Foucault_fees}
\bibfield{author}{\bibinfo{person}{Thierry Foucault}, \bibinfo{person}{Ohad
  Kadan}, {and} \bibinfo{person}{Eugene Kandel}.}
  \bibinfo{year}{2009}\natexlab{}.
\newblock \bibinfo{booktitle}{\emph{{Liquidity cycles and make/take fees in
  electronic markets}}}.
\newblock \bibinfo{type}{{T}echnical {R}eport}.
\newblock


\bibitem[\protect\citeauthoryear{Gini}{Gini}{1936}]%
        {gini}
\bibfield{author}{\bibinfo{person}{Corrado Gini}.}
  \bibinfo{year}{1936}\natexlab{}.
\newblock \showarticletitle{On the measure of concentration with special
  reference to income and statistics}.
\newblock \bibinfo{journal}{\emph{Colorado College Publication, General
  Series}} \bibinfo{volume}{208}, \bibinfo{number}{1} (\bibinfo{year}{1936}),
  \bibinfo{pages}{73--79}.
\newblock


\bibitem[\protect\citeauthoryear{Gladyshev}{Gladyshev}{1965}]%
        {gladyshev1965stochastic}
\bibfield{author}{\bibinfo{person}{EG Gladyshev}.}
  \bibinfo{year}{1965}\natexlab{}.
\newblock \showarticletitle{On stochastic approximation}.
\newblock \bibinfo{journal}{\emph{Theory of Probability \& Its Applications}}
  \bibinfo{volume}{10}, \bibinfo{number}{2} (\bibinfo{year}{1965}),
  \bibinfo{pages}{275--278}.
\newblock


\bibitem[\protect\citeauthoryear{Gould, Porter, Williams, McDonald, Fenn, and
  Howison}{Gould et~al\mbox{.}}{2013}]%
        {gould2013limit}
\bibfield{author}{\bibinfo{person}{Martin~D Gould}, \bibinfo{person}{Mason~A
  Porter}, \bibinfo{person}{Stacy Williams}, \bibinfo{person}{Mark McDonald},
  \bibinfo{person}{Daniel~J Fenn}, {and} \bibinfo{person}{Sam~D Howison}.}
  \bibinfo{year}{2013}\natexlab{}.
\newblock \showarticletitle{Limit order books}.
\newblock \bibinfo{journal}{\emph{Quantitative Finance}} \bibinfo{volume}{13},
  \bibinfo{number}{11} (\bibinfo{year}{2013}), \bibinfo{pages}{1709--1742}.
\newblock


\bibitem[\protect\citeauthoryear{Kirilenko, Kyle, SAMADI, and Tuzun}{Kirilenko
  et~al\mbox{.}}{2017}]%
        {kirilenko_flash_crash}
\bibfield{author}{\bibinfo{person}{Andrei Kirilenko}, \bibinfo{person}{Albert
  Kyle}, \bibinfo{person}{MEHRDAD SAMADI}, {and} \bibinfo{person}{Tugkan
  Tuzun}.} \bibinfo{year}{2017}\natexlab{}.
\newblock \showarticletitle{The Flash Crash: High Frequency Trading in an
  Electronic Market}.
\newblock \bibinfo{journal}{\emph{The Journal of Finance}} (\bibinfo{date}{06}
  \bibinfo{year}{2017}).
\newblock


\bibitem[\protect\citeauthoryear{Kirilenko and Lo}{Kirilenko and Lo}{2013}]%
        {kirilenko_lo}
\bibfield{author}{\bibinfo{person}{Andrei Kirilenko} {and}
  \bibinfo{person}{Andrew Lo}.} \bibinfo{year}{2013}\natexlab{}.
\newblock \showarticletitle{Moore's Law vs. Murphy's Law: Algorithmic Trading
  and Its Discontents}.
\newblock \bibinfo{journal}{\emph{Journal of Economic Perspectives}}
  (\bibinfo{date}{03} \bibinfo{year}{2013}).
\newblock


\bibitem[\protect\citeauthoryear{Kyle}{Kyle}{1985}]%
        {kyle1985continuous}
\bibfield{author}{\bibinfo{person}{Albert~S Kyle}.}
  \bibinfo{year}{1985}\natexlab{}.
\newblock \showarticletitle{Continuous auctions and insider trading}.
\newblock \bibinfo{journal}{\emph{Econometrica: Journal of the Econometric
  Society}} (\bibinfo{year}{1985}), \bibinfo{pages}{1315--1335}.
\newblock


\bibitem[\protect\citeauthoryear{Lan, Kao, Chiang, and Sabharwal}{Lan
  et~al\mbox{.}}{2010}]%
        {FairnessMetric}
\bibfield{author}{\bibinfo{person}{Tian Lan}, \bibinfo{person}{David Kao},
  \bibinfo{person}{Mung Chiang}, {and} \bibinfo{person}{Ashutosh Sabharwal}.}
  \bibinfo{year}{2010}\natexlab{}.
\newblock \showarticletitle{An Axiomatic Theory of Fairness in Network Resource
  Allocation}. In \bibinfo{booktitle}{\emph{Proceedings of the 29th Conference
  on Information Communications}} \emph{(\bibinfo{series}{INFOCOM'10})}.
\newblock


\bibitem[\protect\citeauthoryear{Moulin}{Moulin}{2003}]%
        {hmoulin}
\bibfield{author}{\bibinfo{person}{Hervé Moulin}.}
  \bibinfo{year}{2003}\natexlab{}.
\newblock \bibinfo{booktitle}{\emph{{Fair Division and Collective Welfare}}}.
\newblock \bibinfo{publisher}{The MIT Press}.
\newblock


\bibitem[\protect\citeauthoryear{Nojima, Kojima, and Kubota}{Nojima
  et~al\mbox{.}}{2003}]%
        {morlrobotics}
\bibfield{author}{\bibinfo{person}{Y. Nojima}, \bibinfo{person}{F. Kojima},
  {and} \bibinfo{person}{N. Kubota}.} \bibinfo{year}{2003}\natexlab{}.
\newblock \showarticletitle{Local episode-based learning of multi-objective
  behavior coordination for a mobile robot in dynamic environments}. In
  \bibinfo{booktitle}{\emph{The 12th IEEE International Conference on Fuzzy
  Systems, 2003. FUZZ '03.}}, Vol.~\bibinfo{volume}{1}.
  \bibinfo{pages}{307--312 vol.1}.
\newblock
\urldef\tempurl%
\url{https://doi.org/10.1109/FUZZ.2003.1209380}
\showDOI{\tempurl}


\bibitem[\protect\citeauthoryear{RAWLS}{RAWLS}{1999}]%
        {johnrawls}
\bibfield{author}{\bibinfo{person}{JOHN RAWLS}.}
  \bibinfo{year}{1999}\natexlab{}.
\newblock \bibinfo{booktitle}{\emph{A Theory of Justice}}.
\newblock \bibinfo{publisher}{Harvard University Press}.
\newblock
\showISBNx{9780674000773}
\urldef\tempurl%
\url{http://www.jstor.org/stable/j.ctvkjb25m}
\showURL{%
\tempurl}


\bibitem[\protect\citeauthoryear{Robbins and Monro}{Robbins and Monro}{1951}]%
        {robbins1951stochastic}
\bibfield{author}{\bibinfo{person}{Herbert Robbins} {and}
  \bibinfo{person}{Sutton Monro}.} \bibinfo{year}{1951}\natexlab{}.
\newblock \showarticletitle{A stochastic approximation method}.
\newblock \bibinfo{journal}{\emph{The annals of mathematical statistics}}
  (\bibinfo{year}{1951}), \bibinfo{pages}{400--407}.
\newblock


\bibitem[\protect\citeauthoryear{Roberto}{Roberto}{2016}]%
        {segregation}
\bibfield{author}{\bibinfo{person}{Elizabeth Roberto}.}
  \bibinfo{year}{2016}\natexlab{}.
\newblock \showarticletitle{Measuring inequality and segregation}.
\newblock \bibinfo{journal}{\emph{arXiv preprint arXiv:1508.01167}}
  (\bibinfo{year}{2016}).
\newblock


\bibitem[\protect\citeauthoryear{Roijers, Vamplew, Whiteson, and
  Dazeley}{Roijers et~al\mbox{.}}{2013}]%
        {morlsurvey}
\bibfield{author}{\bibinfo{person}{Diederik~M Roijers}, \bibinfo{person}{Peter
  Vamplew}, \bibinfo{person}{Shimon Whiteson}, {and} \bibinfo{person}{Richard
  Dazeley}.} \bibinfo{year}{2013}\natexlab{}.
\newblock \showarticletitle{A survey of multi-objective sequential
  decision-making}.
\newblock \bibinfo{journal}{\emph{Journal of Artificial Intelligence Research}}
   \bibinfo{volume}{48} (\bibinfo{year}{2013}), \bibinfo{pages}{67--113}.
\newblock


\bibitem[\protect\citeauthoryear{Rosu}{Rosu}{[n.d.]}]%
        {Rosu_adynamic}
\bibfield{author}{\bibinfo{person}{Ioanid Rosu}.}
  \bibinfo{year}{[n.d.]}\natexlab{}.
\newblock \showarticletitle{A dynamic model of the limit order book}.
\newblock \bibinfo{journal}{\emph{Review of Financial Studies}}
  (\bibinfo{year}{[n.\,d.]}), \bibinfo{pages}{2009}.
\newblock


\bibitem[\protect\citeauthoryear{Schwartz}{Schwartz}{2010}]%
        {oregonlaw}
\bibfield{author}{\bibinfo{person}{Jeff Schwartz}.}
  \bibinfo{year}{2010}\natexlab{}.
\newblock \showarticletitle{Fairness, utility and market risk.}
\newblock \bibinfo{journal}{\emph{Oregon Law Review}} (\bibinfo{year}{2010}).
\newblock


\bibitem[\protect\citeauthoryear{Securities and Commission}{Securities and
  Commission}{2012}]%
        {NASDAQ_SEC}
\bibfield{author}{\bibinfo{person}{Securities} {and} \bibinfo{person}{Exchange
  Commission}.} \bibinfo{year}{2012}\natexlab{}.
\newblock \showarticletitle{Order Instituting Proceedings to Determine Whether
  to Approve or Disapprove Proposed Rule Changes Relating to Market Maker
  Incentive Programs for Certain Exchange-Traded Products}.
\newblock  (\bibinfo{year}{2012}).
\newblock


\bibitem[\protect\citeauthoryear{Shelton}{Shelton}{2001}]%
        {morleconomic}
\bibfield{author}{\bibinfo{person}{Christian~Robert Shelton}.}
  \bibinfo{year}{2001}\natexlab{}.
\newblock \showarticletitle{Importance sampling for reinforcement learning with
  multiple objectives}.
\newblock  (\bibinfo{year}{2001}).
\newblock


\bibitem[\protect\citeauthoryear{Speicher, Heidari, Grgic-Hlaca, Gummadi,
  Singla, Weller, and Zafar}{Speicher et~al\mbox{.}}{2018}]%
        {inequalityindices}
\bibfield{author}{\bibinfo{person}{Till Speicher}, \bibinfo{person}{Hoda
  Heidari}, \bibinfo{person}{Nina Grgic-Hlaca}, \bibinfo{person}{Krishna~P
  Gummadi}, \bibinfo{person}{Adish Singla}, \bibinfo{person}{Adrian Weller},
  {and} \bibinfo{person}{Muhammad~Bilal Zafar}.}
  \bibinfo{year}{2018}\natexlab{}.
\newblock \showarticletitle{A unified approach to quantifying algorithmic
  unfairness: Measuring individual \&group unfairness via inequality indices}.
  In \bibinfo{booktitle}{\emph{Proceedings of the 24th ACM SIGKDD International
  Conference on Knowledge Discovery \& Data Mining}}.
  \bibinfo{pages}{2239--2248}.
\newblock


\bibitem[\protect\citeauthoryear{Spooner, Fearnley, Savani, and
  Koukorinis}{Spooner et~al\mbox{.}}{2018}]%
        {SpoonerMarketMaking}
\bibfield{author}{\bibinfo{person}{Thomas Spooner}, \bibinfo{person}{John
  Fearnley}, \bibinfo{person}{Rahul Savani}, {and} \bibinfo{person}{Andreas
  Koukorinis}.} \bibinfo{year}{2018}\natexlab{}.
\newblock \showarticletitle{Market Making via Reinforcement Learning}. In
  \bibinfo{booktitle}{\emph{Proceedings of the 17th International Conference on
  Autonomous Agents and MultiAgent Systems}}.
\newblock


\bibitem[\protect\citeauthoryear{Sutton and Barto}{Sutton and Barto}{2018}]%
        {sutton2018reinforcement}
\bibfield{author}{\bibinfo{person}{Richard~S Sutton} {and}
  \bibinfo{person}{Andrew~G Barto}.} \bibinfo{year}{2018}\natexlab{}.
\newblock \bibinfo{booktitle}{\emph{Reinforcement learning: An introduction}}.
\newblock \bibinfo{publisher}{MIT press}.
\newblock


\bibitem[\protect\citeauthoryear{Theil}{Theil}{1967}]%
        {htheil}
\bibfield{author}{\bibinfo{person}{Henri Theil}.}
  \bibinfo{year}{1967}\natexlab{}.
\newblock \bibinfo{booktitle}{\emph{Economics and information theory}}.
\newblock \bibinfo{publisher}{Amsterdam: North-Holland 1967}.
\newblock


\bibitem[\protect\citeauthoryear{{Twu}, {Mostofi}, and {Egerstedt}}{{Twu}
  et~al\mbox{.}}{2014}]%
        {diversity}
\bibfield{author}{\bibinfo{person}{P. {Twu}}, \bibinfo{person}{Y. {Mostofi}},
  {and} \bibinfo{person}{M. {Egerstedt}}.} \bibinfo{year}{2014}\natexlab{}.
\newblock \showarticletitle{A measure of heterogeneity in multi-agent systems}.
  In \bibinfo{booktitle}{\emph{2014 American Control Conference}}.
\newblock


\bibitem[\protect\citeauthoryear{Venkataraman and Waisburd}{Venkataraman and
  Waisburd}{2007}]%
        {mm}
\bibfield{author}{\bibinfo{person}{Kumar Venkataraman} {and}
  \bibinfo{person}{Andrew~C. Waisburd}.} \bibinfo{year}{2007}\natexlab{}.
\newblock \showarticletitle{The Value of the Designated Market Maker}.
\newblock \bibinfo{journal}{\emph{The Journal of Financial and Quantitative
  Analysis}} \bibinfo{volume}{42}, \bibinfo{number}{3} (\bibinfo{year}{2007}),
  \bibinfo{pages}{735--758}.
\newblock
\showISSN{00221090, 17566916}
\urldef\tempurl%
\url{http://www.jstor.org/stable/27647318}
\showURL{%
\tempurl}


\bibitem[\protect\citeauthoryear{Wah, Wright, and Wellman}{Wah
  et~al\mbox{.}}{2017}]%
        {wah2017welfare}
\bibfield{author}{\bibinfo{person}{Elaine Wah}, \bibinfo{person}{Mason Wright},
  {and} \bibinfo{person}{Michael~P Wellman}.} \bibinfo{year}{2017}\natexlab{}.
\newblock \showarticletitle{Welfare effects of market making in continuous
  double auctions}.
\newblock \bibinfo{journal}{\emph{Journal of Artificial Intelligence Research}}
   \bibinfo{volume}{59} (\bibinfo{year}{2017}), \bibinfo{pages}{613--650}.
\newblock


\bibitem[\protect\citeauthoryear{Wang and Wellman}{Wang and Wellman}{2017}]%
        {wang2017spoofing}
\bibfield{author}{\bibinfo{person}{Xintong Wang} {and}
  \bibinfo{person}{Michael~P Wellman}.} \bibinfo{year}{2017}\natexlab{}.
\newblock \showarticletitle{Spoofing the limit order book: An agent-based
  model}. In \bibinfo{booktitle}{\emph{Proceedings of the 16th Conference on
  Autonomous Agents and MultiAgent Systems}}. \bibinfo{pages}{651--659}.
\newblock


\bibitem[\protect\citeauthoryear{Watkins and Dayan}{Watkins and Dayan}{1992}]%
        {watkins1992q}
\bibfield{author}{\bibinfo{person}{Christopher~JCH Watkins} {and}
  \bibinfo{person}{Peter Dayan}.} \bibinfo{year}{1992}\natexlab{}.
\newblock \showarticletitle{Q-learning}.
\newblock \bibinfo{journal}{\emph{Machine learning}} \bibinfo{volume}{8},
  \bibinfo{number}{3-4} (\bibinfo{year}{1992}), \bibinfo{pages}{279--292}.
\newblock


\bibitem[\protect\citeauthoryear{Young}{Young}{1994}]%
        {peytonyoung}
\bibfield{author}{\bibinfo{person}{H.~Peyton Young}.}
  \bibinfo{year}{1994}\natexlab{}.
\newblock \bibinfo{booktitle}{\emph{Equity: In Theory and Practice}}.
\newblock \bibinfo{publisher}{Princeton University Press}.
\newblock
\showISBNx{9780691043197}
\urldef\tempurl%
\url{http://www.jstor.org/stable/j.ctv10crfx7}
\showURL{%
\tempurl}


\end{thebibliography}

\end{document}